\documentstyle[aps,eqsecnum,amsfonts]{revtex}

\newcommand{\CC}{{\Bbb C}}
\newcommand{\RR}{{\Bbb R}}
\newcommand{\ZZ}{{\Bbb Z}}

\newcommand{\CoinX}[1]{C_0^\infty({#1})}

\newcommand{\DD}{{\cal D}}
\newcommand{\HH}{{\cal H}}

\newcommand{\WW}{{\cal W}}
\newcommand{\OO}{{\cal O}}

\newcommand{\supp}{{\rm supp}\,}
\newcommand{\Span}{{\rm span}\,}

\newcommand{\ip}[2]{\langle #1| #2\rangle}
\newcommand{\act}[2]{\langle #1; #2\rangle}

\newcommand{\Am}{A^{(m)}}
\newcommand{\am}{a^{(m)}}
\newcommand{\Hm}{H^{(m)}}
\newcommand{\hm}{h^{(m)}}

\begin{document}
\draft

\title{Quantum inequalities and `quantum interest' as eigenvalue problems}
\author{Christopher J. Fewster\footnote{Electronic address: {\tt cjf3@york.ac.uk}}}
\address{Department of Mathematics, University of York, Heslington,
York YO10 5DD, United Kingdom}
\author{Edward Teo\footnote{Electronic address: {\tt phyteoe@nus.edu.sg}}}
\address{Department of Applied Mathematics and Theoretical 
Physics, University of Cambridge, Silver Street, 
Cambridge CB3 9EW, United Kingdom\\
Department of Physics, National University of 
Singapore, Singapore 119260}
\date{\today}
\maketitle
\begin{abstract}
Quantum inequalities (QI's) provide lower bounds on the averaged energy
density of a quantum field. We show how the QI's for massless scalar
fields in even dimensional Minkowski space may be reformulated in terms of
the positivity of a certain self-adjoint operator---a generalized
Schr\"odinger operator with the energy density as the potential---and 
hence as an eigenvalue problem. We use this idea to verify that the
energy density produced by a moving mirror in two dimensions is
compatible with the QI's for a large class of mirror trajectories. 
In addition, we apply this viewpoint to
the `quantum interest conjecture' of Ford and Roman, which asserts that
the positive part of an energy density always overcompensates for any
negative components. For various simple models in two and four dimensions
we obtain the best possible bounds on the `quantum interest rate' and on
the maximum delay between a negative pulse and a compensating positive
pulse. Perhaps surprisingly, we find that---in four dimensions---it is
impossible for a positive $\delta$-function pulse of any magnitude to
compensate for a negative $\delta$-function pulse, no matter how close
together they occur.
\end{abstract}
\pacs{04.62.+v 03.70.+k 03.65.Db}

\section{Introduction}

It is well known that the energy density of a quantum field at a given
spacetime point is unbounded from below as a function of the quantum
state~\cite{EGJ}. However, it turns out that the weighted average of the
energy density (either along the worldline of an observer or over a
spacetime region) has a lower bound, provided the weight, or {\em sampling
function\/}, is smooth and has reasonable decay properties. Such bounds are
known as {\em quantum inequalities\/} (QI's), and place severe
constraints on attempts to use quantum fields as a negative energy
source in the construction of, for example, wormholes~\cite{FRWorm}
and `warp drive' spacetimes~\cite{PFWarp}. 

In this paper we will consider QI's on the worldline average 
$\int\rho_\psi(\lambda)f(\lambda)\,d\lambda$, where 
\begin{equation}
\rho_\psi(\lambda):=\left\langle T^{\rm ren}_{ab}(x(\lambda))
\frac{dx^a}{d\lambda}\frac{dx^b}{d\lambda}\right\rangle_\psi
\end{equation}
is the expected energy density in state $\psi$ measured along the
worldline $x(\lambda)$ of an observer, and the sampling function $f$
is smooth and positive. The first inequalities of this type were
obtained by Ford and Roman~\cite{FRqi1,FRqi2} for fields in Minkowski
space in the case where $f$ is the Lorentzian function
$L_\tau(t)=\tau/(\pi(t^2+\tau^2))$. Their results were generalized to
static spacetimes by Pfenning and Ford~\cite{PfF}. General sampling
functions were first treated by Flanagan~\cite{Flan} for massless field
theory in two-dimensional
Minkowski space; more recently, Fewster and Eveson~\cite{FE} used
different methods to obtain QI's valid for general
sampling functions in Minkowski space of arbitrary dimension and fields
of arbitrary mass. This method was extended to static
spacetimes by the present authors~\cite{FT}. Most recently, one of
us has proved a rigorous quantum inequality valid for arbitrary
smooth worldlines in general globally hyperbolic spacetimes and 
arbitrary smooth, compactly supported, positive sampling
functions~\cite{Fqi}. This generalizes and makes precise the results
of~\cite{FE,FT}. Quantum inequalities involving spacetime averages have
been considered by Helfer~\cite{Helf2}, who has established their
existence under similarly general conditions.  

For massless fields in $2m$-dimensional Minkowski space and an observer
with worldline $x(t) = (t,\bbox{0})$, the quantum
inequalities of Flanagan~\cite{Flan} and Fewster and Eveson~\cite{FE} take the form
\begin{equation}
\int \rho_\psi(t) f(t)\,dt \ge -\frac{1}{c_m}\int |(D^m f^{1/2})(t)|^2
\,dt
\label{eq:QI}
\end{equation}
where $D$ is the derivative operator and the $c_m$ are constants given by
\begin{equation}
c_m = \left\{\begin{array}{cl} 6\pi & m=1; \\ 
m\pi^{m-1/2}2^{2m}\Gamma(m-\frac{1}{2}) & m\ge 2. \end{array}
\right.
\end{equation}
The case $m=1$ is due to Flanagan and is tighter than the
corresponding result of Fewster and Eveson which would give
$c_1=4\pi$. Note that the right-hand side of Eq.~(\ref{eq:QI}) is well
defined and finite for a large class of sampling functions, including
all functions equal to the square of a nonnegative Schwartz test
function. 

These inequalities imply that $\rho_\psi$ is non-negative on average.
For example, putting $f(t)=e^{-\alpha t^2}$ in~(\ref{eq:QI}) we obtain
\begin{equation}
\liminf_{\alpha\to 0^+} \int\rho_\psi(t) e^{-\alpha t^2}\,dt \ge 0,
\label{eq:AWEC}
\end{equation}
since the right-hand side of~(\ref{eq:QI}) is of order $\alpha^{m-1/2}$. 
Ford and Roman~\cite{FRcf,FRqi} 
describe this by way of a financial metaphor: negative portions of
$\rho_\psi(t)$ are described as a loan, and positive portions as
repayments. The averaged weak energy condition~(\ref{eq:AWEC})
becomes the statement
that all loans must be repaid in full. However, it is remarkable that 
the positive contributions actually overcompensate for the negative
parts in many examples. The {\em Quantum Interest Conjecture\/} of 
Ford and Roman~\cite{FRcf,FRqi} asserts that this is a general phenomenon:
quantum `loans' must be repaid at a positive rate of interest; moreover,
there is a maximum allowed delay between the loan and the repayments. 
In~\cite{FRqi} they showed that this is true for
the case of a linear combination of two $\delta$-function
pulses\footnote{Energy densities of this form are produced in two
dimensions by moving mirrors with piecewise constant proper
acceleration~\cite{FRqi,FD}.} in two and four dimensions, and also 
obtained lower bounds on the interest rate and upper bounds on the
maximum allowed delay between a negative $\delta$-function pulse and its
compensating positive one (the so-called term of the loan). 
Pretorius~\cite{Pret} has also shown the existence of quantum interest and
maximum delays for much more general energy distributions. His work, and
that of Ford and Roman, relies on optimizing the appropriate 
quantum inequality over a one-parameter family of scaled sampling
functions, and is not guaranteed to give the best possible bounds. 

In this paper we present a new viewpoint on quantum inequalities, 
which yields the best possible results on quantum interest using the
currently available QI's. These results strengthen those of~\cite{FRqi,Pret}. We will study massless fields on
$2m$-dimensional Minkowski space, for which the quantum inequality
\begin{equation}
\int \rho_\psi(t)|g(t)|^2\,dt \ge -\frac{1}{c_m}\int |D^m g(t)|^2\,dt
\label{eq:QI2}
\end{equation}
holds for arbitrary smooth, compactly supported complex-valued functions
$g\in \CoinX{\RR}$ and any Hadamard state $\psi$. For $m\ge 2$, this form
of the quantum inequality is a special case of the general result proved
in~\cite{Fqi} (and also follows from
the derivation given in~\cite{FT} applied to the real and imaginary
parts of $g$ separately); in the case $m=1$ (where
the methods of~\cite{FE,FT,Fqi} give a weaker bound than that
of~\cite{Flan}) the inequality is a consequence of Flanagan's
result.\footnote{Suppose $g\in\CoinX{\RR}$. Then
$f(t)=|g(t)|^2$ is smooth, compactly supported, and pointwise
non-negative, so one may apply Flanagan's bound to $f$, obtaining a right-hand
side equal to $-c_1^{-1}\int (|g|')^2\,dt$. Now $|g|$ is a function of {\em
bounded variation\/}~\cite{RN} 
because $\sum_{k=1}^N |\,|g(t_k)|-|g(t_{k-1})|\,|\le
\int |g'(t)|\,dt$ for any finite 
partition $t_0<t_1<\cdots< t_N$ of some closed
interval containing the support of $g$. Accordingly $|g|$ has a finite
derivative almost everywhere (see \S 4 in~\cite{RN}), and since
$(|g|')^2\le |g'|^2$ at each such point the inequality~(\ref{eq:QI2})
holds.} 

Given a candidate energy density $\rho$, we wish to determine whether or
not $\rho$ is the energy density of a physical quantum state. More
generally, we may ask whether there exists a state whose energy density
agrees with $\rho$ on some (not necessarily bounded) interval
$I\subset\RR$ of the observer's worldline. The quantum inequality
provides an obvious necessary condition: we will say that $\rho$ is 
{\em QI-compatible on $I$\/} with massless scalar fields in $2m$-dimensions
if Eq.~(\ref{eq:QI2}) holds (with $\rho_\psi$ replaced
by $\rho$) for all $g\in\CoinX{I}$. Integrating by parts, this condition
becomes the inequality
\begin{equation}
\ip{g}{\Hm g} \ge 0,\qquad \forall g\in\CoinX{I}
\end{equation}
where $\ip{\,\cdot\,}{\,\cdot\,}$ is the usual $L^2$-inner product and $\Hm$ is
the differential operator
\begin{equation}
\Hm = (-1)^m D^{2m} + c_m\rho. \label{eq:Hform}
\end{equation}
At a formal level, then, the QI-compatibility of $\rho$ on $I$ is
equivalent to the positivity of $\Hm$ as an operator on $L^2(I)$.
Accordingly, the problem is reduced to finding the infimum of the
spectrum $\sigma(\Hm)$ of $\Hm$, and hence to an eigenvalue problem with 
$\rho$ as a potential. 

Of course, we have skated over various technical issues here, notably
the conditions on $\rho$ necessary for $\Hm$ to exist 
a self-adjoint operator on $L^2(I)$ (especially
if $\rho$ is allowed to be distributional\footnote{Although the usual 
(Hadamard) class of physically admissible states yields a smooth energy
density, we relax the condition here to allow discussion of, for
example, moving mirror models with nonsmooth proper
acceleration~\cite{FD,FV}.}), the possibility that 
$\inf\sigma(\Hm)$ might be a point of the continuous or singular
continuous spectrum, rather than an eigenvalue, and the question of what
boundary conditions are appropriate at $\partial I$ (if this is
nonempty). These issues are addressed in Sects.~\ref{sect:techapp}
and~\ref{sect:techres}, where we describe classes $\WW_m$ of
candidate energy densities for which the arguments above
may be made rigorous. The $\WW_m$ are constructed from certain 
distributional Sobolev spaces, and  
include progressively more singular
distributions as $m$ increases: for example, each $\WW_m$ 
contains the $\delta$-distribution and its first $m-1$ derivatives.

For any $\rho\in\WW_m$, we will define $\Hm$ as a self-adjoint 
operator on $L^2(I)$ by first regarding the 
expression~(\ref{eq:Hform}) as a sum of
quadratic forms. We will then show that the following are equivalent:
\begin{enumerate}
\item $\rho\in\WW_m$ is QI-compatible on $I$ with massless fields in
$2m$-dimensions; 
\item The operator $\Hm$ on $L^2(I)$ is positive;
\item $\Hm$ has no strictly negative {\em eigenvalues\/}. 
\end{enumerate}
The appropriate boundary conditions for the
eigenvalue problem require the eigenfunction and its
first $m-1$ derivatives to vanish on $\partial I$. 

In fact, the reader who does not wish to be burdened with the details
could skip most of Sects.~\ref{sect:techapp} and~\ref{sect:techres},
apart from the definition of $\WW_m$ and the statements of
Theorems~3.2 and~3.3. However, these sections 
may be of independent technical interest, as it appears that the
quadratic form technique for defining 
Hamiltonians with singular potentials has
previously been applied only to potentials which are locally integrable
functions~\cite{Simon,SchPap} (using techniques adapted to this
case) or to particular distributions given explicitly as separable
interactions (see, e.g.,~\cite{GHM}, Appendix~G of~\cite{AGHH} or
Example~3 in Section X.2 of~\cite{RSii}). This is largely because
the technique has somewhat limited utility for the Schr\"odinger
equation in $d\ge 2$ dimensions, in which case even the $\delta$-function
is too singular to be treated.\footnote{The $\delta$-distribution is
usually treated in terms of self-adjoint extensions (see,
e.g.~\cite{AGHH}) in $d\le 3$ dimensions, while more singular
distributions may be treated on enlarged, possibly indefinite, 
inner product spaces (see~\cite{vDT,Few} and 
references therein).} In the present setting, however, the spacetime
dimension of the original field theory determines the order of
the unperturbed operator $(-1)^m D^{2m}$, and the classes $\WW_m$ 
become more general with increasing $m$ as mentioned above.

As well as reformulating the quantum inequalities,
the eigenvalue viewpoint allows us to demonstrate the existence
of maximal pulse separations under very general conditions
(Theorem~3.3) albeit in a rather nonconstructive fashion.
In addition, one may gain more insight into the quantum interest
conjecture by analogy with quantum mechanics on the line. Negative
energy loans become potential wells; repayments become potential
barriers. To be QI-compatible, the barrier must be sufficiently `large'
in some sense to bounce a particle out of the potential well and prevent
it being bound. Further comments in this direction are presented in the
conclusion. 

However, the main thrust of this paper is that our viewpoint provides
a practical method for determining QI-compatibility of candidate energy
densities, and for obtaining the best possible bounds on maximal pulse
separations and quantum interest rates. This is illustrated by various
examples, starting in Sect.~\ref{sect:mirror} with a
simple argument to verify the QI-compatibility of the energy densities
arising from a large class of moving mirror trajectories in two
dimensions. In Sect.~\ref{sect:del2d}, we discuss candidate energy
densities consisting of isolated $\delta$-function pulses in two
dimensions and obtain
sharp bounds on the maximal separation between a $\delta$-function loan
and repayments (which need not be of $\delta$-function form). For the
case where the repayment is also a $\delta$-function pulse, we also 
determine the minimum quantum interest rate, thereby sharpening the
bounds obtained by Ford and Roman~\cite{FRqi}. We show that the quantum
interest rate becomes unboundedly large as the delay between loan and
repayment approaches its maximal (finite) value. By contrast,
the lower bound on the interest rate obtained in~\cite{FRqi}
remains finite in this limit, although this does not contradict
our results. 

Section~\ref{sect:del4d} repeats this analysis in four dimensions. 
Surprisingly, in this case it turns out that no positive
$\delta$-function pulse of any magnitude can compensate for a 
negative $\delta$-function pulse, no matter how close together they occur.
Thus all $\delta$-function combinations of
the type considered by Ford and Roman violate the
quantum inequality in four dimensions, illustrating the added
strength of our method. However, we emphasize that a single negative
$\delta$-function pulse can be part of a QI-compatible energy density.
To this end, Section~\ref{sect:delstep} considers a negative
$\delta$-function pulse followed at some later time by a constant positive
energy density of finite magnitude but infinite duration. Again, it turns
out that the magnitude of the compensating pulse must become unboundedly large as
the maximal delay is approached. 

Our results are all derived for massless fields in even spacetime
dimensions. However, as Pretorius~\cite{Pret} has
emphasized, the quantum inequalities for massive fields are stronger
than those for massless fields (except in two dimensions)
and so our results also provide bounds on the massive case. One could
interpret the massive QI's directly in terms of a positivity condition,
but the analog of $H$ would be a pseudodifferential operator which
would be much harder to analyze. Nonetheless, this may prove to be a
fruitful viewpoint in deepening our understanding of quantum
inequalities. 
 
\section{Technical apparatus} \label{sect:techapp}

In this section we set up the technical framework in which our main
results are proved. 

\subsection{Sobolev spaces}

For completeness and to fix the notation, we briefly describe 
the various Sobolev spaces which appear in the sequel, mainly
following~\cite{Adams}. 

First, let $\Omega$ be any open subset of $\RR$, $m=0,1,2,\ldots$ and 
$1\le p\le\infty$. The spaces $W^{m,p}(\Omega)$ and $W_0^{m,p}(\Omega)$ are
defined as follows: $W^{m,p}(\Omega)$ is 
the space of functions in $L^p(\Omega)$ whose first $m$ weak derivatives
are also in $L^p(\Omega)$, equipped with the norm
\begin{eqnarray}
\|u\|_{m,p} & =& 
\left(\sum_{r=0}^m \|D^ru\|_p^p\right)^{1/p}, \qquad 1\le p<\infty, \nonumber\\
\|u\|_{m,\infty}& =& \max_{0\le r\le m} \|D^r u\|_\infty,
\end{eqnarray}
while $W_0^{m,p}(\Omega)$ is defined to be the closure of
$\CoinX{\Omega}$ in $W^{m,p}(\Omega)$. Here,
$\|\cdot\|_p$ denotes the usual $L^p$-norm on $\Omega$,
\begin{eqnarray}
\|\varphi\|_p &=&\left(\int_\Omega |\varphi(t)|^p\,dt\right)^{1/p}, \qquad 1\le
p<\infty, \nonumber\\
\|\varphi\|_\infty &=&\sup_{t\in\Omega} |\varphi(t)|.
\end{eqnarray}
In general $W_0^{m,p}(\Omega)$ is a proper subspace of
$W^{m,p}(\Omega)$, but we note the special case
$W_0^{m,p}(\RR)=W^{m,p}(\RR)$ ($1\le p<\infty$). 

An important property of these spaces is given by the Sobolev embedding
theorem (Theorem~5.4 in~\cite{Adams}) which we state for the case $p=2$: 

\medskip
{\em Proposition 2.1\/}. Let $I$ be an open interval of $\RR$. 
For $m\ge 1$, elements of $W^{m,2}(I)$ may be identified with 
$C^{m-1}$-functions on the closure of $I$, and the identification map is
continuous. In consequence, elements of $W_0^{m,2}(I)$ vanish along with
their first $m-1$ derivatives on the boundary of $I$. 
\medskip

For $m\ge 0$ and $1\le p<\infty$, the dual space of $W^{m,p}(\RR)$ is
denoted $W^{-m,p'}(\RR)$, where the conjugate index $p'$ is defined by
\begin{equation}
p'=\left\{\begin{array}{cl} p/(p-1) & 1<p<\infty; \\ \infty & p=1.
\end{array}\right.
\end{equation} 
This space may also be characterized as the set of distributions in
$\DD'(\RR)$ of the form $\omega=\sum_{r=0}^m D^r\psi_r$, where each
$\psi_r$  belongs to $L^{p'}(\RR)$. Distributions of this form act
antilinearly on $W^{m,p}(\RR)$ by
\begin{equation}
\omega:f\mapsto \act{f}{\omega}=\sum_{r=0}^m \int
(-1)^r\psi_r(t)\overline{D^r f(t)}\,dt,
\end{equation}
and the norm $\|\cdot\|_{-m,p'}$ is defined to be the operator norm of this
map, namely, 
\begin{equation}
\|\omega\|_{-m,p'} = 
\sup_{\scriptstyle f\in W^{m,p}(\RR)\atop \scriptstyle f\not =0} 
\frac{|\act{f}{\omega}|}{\|f\|_{m,p}}.
\end{equation}

Finally, for any $m\in\ZZ$ the local Sobolev space $W^{m,2}_{\rm
loc}(\Omega)$ is defined to be the space of $u\in \DD'(\Omega)$ 
such that $\chi u$ is the restriction to $\Omega$ of
an element of $W^{m,2}(\RR)$ for all $\chi\in\CoinX{\Omega}$. These have the
following property:

\medskip
{\em Proposition 2.2\/}. If $f$ and $D^{2r}f$ both belong to $W^{m,2}_{\rm loc}(\Omega)$ for some
integer $r\ge 1$ then $f\in W^{m+2r,2}_{\rm loc}(\Omega)$. 
\medskip

See Section~IX.6 in~\cite{RSii} (or Theorem~11.1.8 in~\cite{Hor2} for a
very general setting).

\subsection{Quadratic forms}

As mentioned in the introduction, the self-adjoint operator $\Hm$ is
defined as a sum of quadratic forms. We briefly review the theory 
of quadratic forms in Hilbert space~\cite{RSi,Kato,SchSpec}
and then prove two technical results
which will underpin our discussion in Sect.~\ref{sect:techres}. 

Let $\HH$ be a separable complex Hilbert space with inner product
$\ip{\,\cdot\,}{\,\cdot\,}$. A {\it quadratic form} on $\HH$ is a map
$a:Q(a)\times Q(a)\to \CC$, where $Q(a)$ is a dense linear subset of
$\HH$ called the {\em form domain\/}, with $a$
conjugate linear in the first slot and linear in the second. The form is
said to be {\em semi-bounded\/} if there exists
$M\ge 0$ so that $a(\varphi,\varphi)\ge -M\ip{\varphi}{\varphi}$ for
all $\varphi\in Q(a)$, and {\em positive\/} if we may take $M=0$. 
A semi-bounded form on a complex Hilbert space is automatically
{\em symmetric\/}, i.e., $\overline{a(\varphi,\psi)}=a(\psi,\varphi)$
for all $\varphi,\psi\in Q(a)$. 

If $a$ is semi-bounded from below by $-M$, it determines an inner product
$\ip{\,\cdot\,}{\,\cdot\,}_{+1}$ on $Q(a)$ by
\begin{equation}
\ip{\varphi}{\psi}_{+1}=a(\varphi,\psi)+(M+1)\ip{\varphi}{\psi}.
\label{eq:ip}
\end{equation}
If $Q(a)$ is complete with respect to the corresponding norm
$\|\cdot\|_{+1}$, the form $a$ is said to be {\em closed\/} and $Q(a)$
becomes a Hilbert space, sometimes denoted $\HH_{+1}$. It is easily seen
that $\psi_n\to\psi$ in $\HH_{+1}$ if and only if $\psi_n\to\psi$ in
$\HH$ and $a(\psi_n,\psi_n)\to a(\psi,\psi)$. 
In consequence, if $\DD$ is dense in $Q(a)$ in the $\|\cdot\|_{+1}$
norm, i.e., if $\DD$ is a {\em form core\/} for $a$, then $a$ agrees with
its continuous extension from $\DD$ to $Q(a)$. A particular case of this
is:

\medskip
{\em Proposition 2.3\/}.
If a semi-bounded closed form $a$ is positive on a form
core $\DD$ (i.e., $a(\psi,\psi)\ge 0$ for all $\psi\in\DD$) then it is
positive.  
\medskip

{\em Proof.\/} For each $\psi\in Q(a)$ there exists a sequence of
elements $\psi_n\in\DD$ with $\psi_n\to\psi$ in $\|\cdot\|_{+1}$.
Thus $a(\psi,\psi) = \lim_{n\to\infty} a(\psi_n,\psi_n)\ge 0$.
$\square$
\medskip

Each semi-bounded form $a$ may be associated uniquely with a
self-adjoint operator $A$ whose domain $D(A)$ is contained in $Q(a)$ and
such that 
\begin{equation}
a(\varphi,\psi)=\ip{\varphi}{A\psi} 
\label{eq:aA}
\end{equation}
holds for all $\varphi\in Q(a)$
and $\psi\in D(A)$ (Theorem~VI-2.1 of~\cite{Kato}). This operator is
specified as follows: if $\varphi\in Q(a)$, $\psi\in\HH$, and 
$a(\chi,\varphi)=\ip{\chi}{\psi}$ for all $\chi$ belonging to a
form core for $a$, then $\varphi\in D(A)$ and $A\varphi=\psi$. 
In addition, $D(A)$ is itself a form core for $a$. 

On the other hand,
if $A$ is a semi-bounded symmetric operator\footnote{An operator 
$A$ is {\em semi-bounded\/} if there is an $M\ge 0$ such that 
$\ip{\psi}{A\psi}\ge -M\|\psi\|^2$ for all $\psi\in D(A)$ and
{\em symmetric\/} if
$\ip{\varphi}{A\psi}=\ip{A\varphi}{\psi}$ for all $\varphi,\psi\in 
D(A)$---note that this is {\em not\/} imply that $A$ is self-adjoint (see
Ch.~VIII in~\cite{RSi}).} on $\HH$,  
there is a unique closed semi-bounded quadratic form
$a$ such that $D(A)$ is a form core for $a$ and Eq.~(\ref{eq:aA}) holds
for all $\varphi,\psi\in D(A)$ (Theorem X.23 in~\cite{RSii}). In turn,
$a$ is associated with a self-adjoint operator $\widehat{A}$ called 
the {\em Friedrichs extension\/} of $A$; it is the unique self-adjoint
extension whose domain is contained in $Q(a)$, and the lower bound of its
spectrum is equal to the lower bounds of $a$ and $A$. 

Applying this result in the case where $A$ is itself self-adjoint, we
establish a bijection between semi-bounded self-adjoint
operators and closed semi-bounded quadratic forms (of course, in this
case $\widehat{A}=A$). For this reason, 
the distinction between self-adjoint operators and
the corresponding quadratic forms is often blurred, and we speak of 
the form core of an operator or write $Q(A)$ for the form domain of the
quadratic form associated with $A$. 

To illustrate these ideas, let $I\subset\RR$ be an open (not necessarily
bounded) interval, and consider the operator $(-1)^m D^{2m}$ on
$\CoinX{I}\subset L^2(I)$, which is easily seen to be symmetric and 
positive. The form domain of its Friedrichs
extension $\Am_I$ is therefore the closure of 
$\CoinX{I}$ in the form norm $\|\cdot\|_{+1}^2
= \|D^m \cdot\|_2^2+\|\cdot\|_2^2$. But this norm is equivalent to the 
$\|\cdot\|_{m,2}$ Sobolev norm by Corollary 4.16 in~\cite{Adams}, so  
$Q(\Am_I)=W_0^{m,2}(I)$. Accordingly the elements of $Q(\Am_I)$ are
$C^{m-1}$ functions which vanish along with their first $m-1$
derivatives on $\partial I$; that is, they obey Dirichlet boundary
conditions. The domain $D(\Am_I)$ is found as follows: we have
$\varphi\in D(\Am_I)$ if and only if $\varphi\in W_0^{m,2}(I)$ and, 
for some $\psi\in L^2(I)$ we have $\ip{\varphi}{(-1)^m D^{2m}
f}=\ip{\psi}{f}$ for all $f\in\CoinX{I}$. This holds if and only if the 
$2m$'th weak derivative of $\varphi$ belongs to $L^2(I)$. We conclude that
$D(\Am_I) = W_0^{m,2}(I)\cap W^{2m,2}(I)$. 

We now turn to the definition of operators as sums of quadratic forms. 
For simplicity, assume $A$ is a positive self-adjoint
operator with associated form $a$. Suppose $b$ is a symmetric quadratic
form defined on $Q(A)$. We say that $b$ is {\em $A$-form bounded\/} 
(or form bounded
relative to $A$) if there exist real constants $\alpha,\beta$ such that
\begin{equation}
|b(\varphi,\varphi)|\le \alpha a(\varphi,\varphi) + \beta
\ip{\varphi}{\varphi}, 
\qquad \forall \varphi\in Q(A).
\label{eq:rfb}
\end{equation}
The {\em relative form bound\/} of $b$ with respect to $A$ is the infimum
of the set of $\alpha$ for which Eq.~(\ref{eq:rfb}) holds for some
choice of $\beta$. By the Riesz lemma, if $b$ is $A$-form bounded then
there is a bounded self-adjoint operator $B:\HH_{+1}\to\HH_{+1}$ such that
\begin{equation}
b(\varphi,\psi) = \ip{\varphi}{B\psi}_{+1}, \qquad \forall
\varphi,\psi\in Q(A).
\end{equation}
If $B$ is compact, then $b$
is said to be {\em $A$-form compact\/}; furthermore, the relative form
bound may be shown to vanish in this case (cf.~problem~39 in Ch.~XIII
of~\cite{RSiv}). Thus, for each $\alpha>0$ there exists $\beta$ such
that~(\ref{eq:rfb}) holds. 

Our results below will be applications of the following general result.
Here, the {\em essential spectrum\/} $\sigma_{\rm ess}(A)$ of a
self-adjoint operator $A$ is 
equal the set of all spectral points of $A$ other than
isolated eigenvalues of finite multiplicity. 

\medskip
{\em Theorem 2.4\/}.
Let $A$ be a positive self-adjoint operator and let $b$ and $c$ be symmetric 
quadratic forms defined on $Q(A)$. Suppose that $b$ is positive and $A$-form
bounded, and $c$ is $A$-form compact. Then 
$h=a+b+c$ (with form domain $Q(A)$) 
is the quadratic form of a unique semi-bounded self-adjoint
operator $H$ with $Q(H)=Q(A)$. Any form core of $A$ is a form core for
$H$ and $\inf\sigma_{\rm ess}(H)\ge\inf \sigma_{\rm ess}(A)$. In
particular, $\sigma_{\rm ess}(H)\subset[0,\infty)$. 
\medskip

{\em Proof.\/}
By hypothesis on $b$ and $c$
there exist $\alpha,\beta,\gamma>0$ such that
\begin{equation}
0\le b(\psi,\psi) \le \alpha a(\psi,\psi) + \beta\|\psi\|^2
\end{equation}
and
\begin{equation}
|c(\psi,\psi)| \le \frac{1}{2} a(\psi,\psi) + \gamma\|\psi\|^2
\end{equation}
for all $\psi\in\HH_{+1}$. Thus
\begin{equation}
\frac{1}{2}\|\psi\|_{+1}^2 \le \|\psi\|_{+1,h}^2 \le \kappa
\|\psi\|_{+1}^2
\end{equation}
for some $\kappa>1/2$, where
\begin{equation}
\|\psi\|_{+1,h}^2= h(\psi,\psi) + (\gamma+1)\|\psi\|^2
\end{equation}
and we see that $\|\cdot\|_{+1,h}$
and $\|\cdot\|_{+1}$ are equivalent norms. Accordingly, $h$ is a closed
semi-bounded quadratic form on $Q(A)$ and is therefore the quadratic
form of a
unique self-adjoint operator $H$ with $Q(H)=Q(A)$ by Theorem~VI-2.1
in~\cite{Kato}. 
The statement concerning form cores
follows because any $\|\cdot\|_{+1}$-dense subset of $Q(A)$ is also
$\|\cdot\|_{+1,h}$-dense. Exactly the same argument shows that $h'=a+c$
is the quadratic form of a unique self-adjoint $H'$ with
$Q(H')=Q(H)=Q(A)$. 

To show that $\inf\sigma_{\rm ess}(H)\ge \inf\sigma_{\rm ess}(A)$,
we need only to consider the case where $H$ has a
nonempty essential spectrum. Since $c$ is $A$-form compact we have
$\sigma_{\rm ess}(H')=\sigma_{\rm ess}(A)$ by a generalization of
Weyl's theorem\footnote{See, for example,
problem~39 in Ch.~XIII of~\cite{RSiv}. This may be proved using an
argument similar to that of Theorem~6.11 in Ch.~1 of~\cite{SchSpec}.}
so it is enough to show that $\inf\sigma_{\rm ess}(H)\ge 
\inf\sigma_{\rm ess}(H')$.  
Accordingly, for $n=1,2,\ldots$ we define
\begin{equation}
\mu_n = \sup_{\varphi_1,\ldots,\varphi_{n-1}}
\inf_{
\scriptstyle \psi\perp\Span\{\varphi_1,\ldots,\varphi_{n-1}\}\atop
\scriptstyle \|\psi\|=1;~\psi\in Q(H)} h(\psi,\psi)
\end{equation}
and $\mu_n'$ by the same formula with $H$ and $h$ replaced by $H'$ and
$h'$. Positivity of $b$ (and the fact that $Q(H)=Q(H')=Q(A)$)
entails that $\mu_n\ge \mu_n'$ for each $n$. 
By the min-max principle (Theorems XIII.1 and XIII.2 in~\cite{RSiv}) 
the $\mu_n$ are nondecreasing and tend
to $\inf\sigma_{\rm ess}(H)$ as $n\to\infty$. Hence the $\mu_n'$ form a
bounded monotonic sequence which therefore converges to 
$\inf\sigma_{\rm ess}(H')$, thus yielding the
required inequality. $\square$ 
\medskip

We remark that the construction of $H$ is based directly on the proof of
the KLMN theorem~\cite{RSii} using some ideas drawn from~\cite{SchPap}.

\medskip
{\em Corollary 2.5\/}.
The following are equivalent: 
(i)~$h$ is positive on a form core for $A$; 
(ii)~$H$ is positive; 
(iii)~$H$ has no strictly negative eigenvalues. 
\medskip

{\em Proof.\/} Statements~(i) and~(ii) are equivalent by
Proposition~2.3; statements~(ii) and~(iii) are equivalent
because any negative
spectral points of $H$ are necessarily isolated eigenvalues since 
$\sigma_{\rm ess}(H)\subset [0,\infty)$. 
$\square$
\medskip

\section{Main technical results} \label{sect:techres}

We now apply the foregoing results to the theory of quantum
inequalities. Let $I$ be any open (not necessarily
bounded) interval in $\RR$ and, as in Sect.~\ref{sect:techapp},
let $\Am_I$ be the Friedrichs extension of the positive
symmetric operator $(-1)^m D^{2m}$ on $\CoinX{I}\subset L^2(I)$.
Each $\rho\in W^{-m,\infty}(\RR)$ determines a quadratic form
$\rho_I$ on $Q(\Am_I)$ by the formula
\begin{equation}
\rho_I(f,g)=\act{(\jmath f)\overline{\jmath g}}{\rho} 
\label{eq:rhoIdef}
\end{equation}
and $\rho_I$ is symmetric if $\rho$ is {\em real\/} in the sense that
$\overline{\act{f}{\rho}}=\act{\overline{f}}{\rho}$ for all $f\in
W^{m,1}(\RR)$.
Here, $\jmath f$ is the element of $W^{m,2}(\RR)$ agreeing with $f$ on
$I$ and vanishing elsewhere.\footnote{The extension map $\jmath$ is a
continuous isometry of $W_0^{m,2}(I)$ into $W^{m,2}(\RR)$ 
(Lemma 3.22 in~\cite{Adams}). Formula~(\ref{eq:rhoIdef}) makes sense
because the product of two $W^{m,2}$-functions is in $W^{m,1}$ 
by Leibniz' rule and H\"older's inequality.}

Comparing with Eq.~(\ref{eq:QI2}), the candidate energy density $\rho$
is QI-compatible on $I$ (with massless scalar fields in $2m$-dimensions)
if and only if the form sum 
$\hm_I=\am_I+c_m\rho_I$ is positive on $\CoinX{I}$, i.e., $\hm_I(g,g)\ge
0$ for all $g\in\CoinX{I}$. To turn this into
spectral information on a self-adjoint operator we restrict the 
class of candidate energy densities so as to satisfy the hypotheses of
Theorem~2.4. 

Accordingly, let $\WW_m$ be the set of
real $\rho\in W^{-m,\infty}(\RR)$ such that $\rho=\rho_1+\rho_2$ with
$\rho_1$ positive (i.e., $\act{f}{\rho_1}\ge 0$ for all pointwise
non-negative test functions $f$) and $\rho_2\in W^{-m,2}(\RR)\cap
W^{-m,\infty}(\RR) + \left(W^{-m,\infty}(\RR)\right)_\epsilon$.\footnote{That
is, for any $\epsilon>0$, $\rho_2$ may be written $\rho_2=\rho_3+\rho_4$ with 
$\rho_3\in W^{-m,2}(\RR)\cap
W^{-m,\infty}(\RR)$, and $\rho_4\in W^{-m,\infty}(\RR)$ with
$\|\rho_4\|_{-m,\infty}<\epsilon$.} These distributional classes are
more than adequate to discuss the examples considered by Ford and
Roman~\cite{FRqi} and in the present paper. Indeed, $\WW_m$ contains the
$\delta$-distribution and its first $m-1$ derivatives, and also includes
(by virtue of the definition of $\rho_2$) 
potentials with very slowly decaying negative tails. 
The key property of these energy densities is given by the
following result, whose proof is given in Appendix~\ref{appx:Pf}: 

\medskip
{\em Lemma 3.1\/}.
Suppose $\rho\in \WW_m$ and write $\rho=\rho_1+\rho_2$ as above. 
Let $\rho_{i,I}$ be the quadratic forms induced on $Q(\Am_I)$ by the
$\rho_i$ according to Eq.~(\ref{eq:rhoIdef}). Then $\rho_{1,I}$ is 
positive and $\Am_I$-form bounded and $\rho_{2,I}$ is $\Am_I$-form compact. 
\medskip

Lemma~3.1 and Theorem~2.4 show that, if $\rho\in\WW_m$,
then $\hm_I$ is the quadratic form of a self-adjoint operator $\Hm_I$ with
quadratic form domain equal to $Q(\Am_I)=W_0^{m,2}(I)$ and $\sigma_{\rm
ess}(\Hm_I) \subset [0,\infty)$. We also
know that $\CoinX{I}$ is a form core for $\Hm_I$. 
Our main technical result now follows immediately from 
Corollary~2.5:

\medskip
{\em Theorem 3.2\/}.
Suppose $\rho\in\WW_m$ and let $I$ be an open (not necessarily bounded)
interval of $\RR$. Define $\Hm_I$ as above.
Then the following are equivalent: (i)~$\rho$ is
QI-compatible on $I$; (ii)~$\Hm_I$ is positive; (iii)~$\Hm_I$ has no
strictly negative eigenvalues. 
\medskip

The operator domain $D(\Hm_I)$ of the operator $\Hm_I$ is the space of $g\in
W_0^{m,2}(I)$ for which the distribution 
$(-1)^m D^{2m}g+\rho g$ may be identified with an element of $L^2(I)$,
which we then denote $\Hm_I g$. Here, $\rho g$ is to be
understood as the distribution acting on $W_0^{m,2}(I)$ by 
\begin{equation}
\rho g:f\mapsto \rho_I(f,g).
\end{equation} 
It follows that $\lambda$ is an eigenvalue of $\Hm_I$ if and only if
there exists $g\in W_0^{m,2}(I)$ obeying the distributional eigenvalue
equation 
\begin{equation}
(-1)^m D^{2m}g + c_m\rho g = \lambda g. \label{eq:eval}
\end{equation}
An important part of this result is that it prescribes  
the regularity and boundary conditions obeyed by 
$g$; in particular, by the embedding result Proposition~2.1,
$g$ must be at least $C^{m-1}$ and
vanish along with its first $m-1$ derivatives on $\partial I$. 
In many cases, however, elliptic regularity arguments allow us to deduce
that $g$ enjoys a greater degree of smoothness. 

As an example, consider $\rho=\sum_{n=1}^N \alpha_n \delta_{\tau_n}$,
where $\delta_{\tau_n}$ is the $\delta$-distribution centered at $\tau_n$, 
and the $\alpha_n$ are real. Suppose also that the $\tau_n$ are 
contained in the open interval $I$. For any $m=1,2,\ldots$ we have 
$\rho\in W^{-m,2}(\RR)\cap W^{-m,\infty}(\RR)$ so QI-compatibility
reduces to the eigenvalue problem~(\ref{eq:eval}). Since each $g\in
W_0^{m,2}(I)$ is continuous, $\rho g = \sum_{n=1}^N \alpha_n g(\tau_n)
\delta_{\tau_n}$ and belongs to $W_{\rm loc}^{-1,2}(I)$. Thus if $g$
solves the eigenvalue problem we have
\begin{equation}
(-1)^m D^{2m}g  = \lambda g -\sum_{n=1}^N \alpha_n g(\tau_n)
\delta_{\tau_n} \in W_{\rm loc}^{-1,2}(I)
\end{equation}
from which it follows that $g\in W^{2m-1,2}_{\rm loc}(I)$ by
Proposition~2.2 since both $g$ and $D^{2m}g$ belong to
$W^{-1,2}_{\rm loc}(I)$. Thus 
$g$ has $2m-2$ continuous derivatives on the closure of
$I$. In general, similar arguments show that $g$ is in fact smooth on 
any open set excluding the singular support of $\rho$.

Theorem~3.2 fulfills our stated aim of reducing
QI-compatibility to
an eigenvalue problem. We will use this viewpoint to obtain quantitative
results in the following sections; first, however, we prove the
existence of maximal pulse separation times in this setting. 

\medskip
{\em Theorem 3.3\/}.
Suppose $\rho\in \WW_m$ has compact support $\supp\rho$, and let
$\OO_\rho$ be the set of open intervals $I$ containing $\supp \rho$
on which $\rho$ is QI-compatible, ordered by inclusion. 
If $\OO_\rho$ is nonempty, it contains at least one maximal element. 
\medskip

{\em Proof.\/} 
By Zorn's lemma, it is enough to show
that every totally ordered subset of $\OO_\rho$ has an upper bound
belonging to $\OO_\rho$. Let $I_\alpha$ be any totally ordered set
of nested open intervals in $\OO_\rho$. By Theorem~3.2
each $\Hm_{I_\alpha}$ is positive. Because the
open interval $I=\cup_\alpha I_\alpha$ is an upper bound for the
$I_\alpha$'s it suffices to show that $I\in\OO_\rho$. Now, 
any $f\in\CoinX{I}$ belongs to $\CoinX{I_\alpha}$
for some (indeed, all sufficiently large) $\alpha$ and so
$\ip{f}{\Hm_If} = \ip{f}{\Hm_{I_\alpha}f}\ge 0$. Hence $\Hm_I$ is
positive on the form core $\CoinX{I}$, and is therefore positive. 
Thus $I\in\OO_\rho$ and the result is proved. $\square$
\medskip

The significance of
this result is that any maximal element of $\OO_\rho$ is a maximal interval in
which a physical energy density $\rho_\psi$ can agree with $\rho$. For
example, if $\rho$ is supported in $[-t_0,t_0]$, and it turns
out that, for some $T>t_0$, $(-\infty,T)$ is a maximal interval in this
sense, then any physical energy density $\rho_\psi$ agreeing with $\rho$
on $(-\infty,T)$ must have a compensating pulse starting no later than
$t=T$. We
note that there may be more than one maximal interval; if, in the above
example, $\rho$ was symmetric about the origin, then $(-T,\infty)$ would
also be a maximal interval. 

To summarize, we have seen that the QI-compatibility of a large class of
distributional candidate energy densities on an open interval $I$ 
may be determined by solving
an eigenvalue problem subject to known regularity
properties in $I$ and boundary conditions on $\partial I$. We have also
established (in an admittedly nonconstructive fashion) the existence of
constraints on `pulse separation' in this setting. The main importance of
these results, to which we now turn, is that they also provide a
practical method of 
verifying QI-compatibility and also obtaining sharp
pulse separation bounds and quantum interest rates in particular classes
of candidate energy densities.

\section{Moving mirrors in two dimensions} \label{sect:mirror}

Suppose an inertial observer in two-dimensional Minkowski space measures
the energy density produced by a perfectly reflecting mirror, which is
initially at rest in the frame of the observer and subsequently
describes a trajectory which is uniform at late times. One would expect
the energy density to be QI-compatible (on $\RR$) for all `reasonable'
trajectories---indeed Davies~\cite{Davies} and (in essence) Ford~\cite{Ford78}
used this assumption as a starting point for the derivation of simple QI
bounds; we will now show how our framework allows a straightforward
verification of this fact for a large class of trajectories. 

We suppose the observer follows a worldline $(t,0)$, and that the
mirror follows the worldline $(t,z(t))$ with $z(t)<0$ for all $t$ to
prevent collisions between the observer and mirror. Similar arguments
would apply if $z$ was everywhere positive. For simplicity, we assume
that $z$ is smooth and that the mirror has boundedly subluminal velocity, 
i.e., $\sup_t |\dot{z}(t)|<1$. The
assumptions on the initial and final states of motion imply that
$z(t)=z_0<0$ for all sufficiently large negative $t$ and that
$\ddot{z}(t)$ vanishes for large positive $t$. In addition, we must have
$\dot{z}\le 0$ at late times to avoid collisions.\footnote{These
conditions are too restrictive to encompass the nonsmooth trajectory
discussed in~\cite{FRqi}, which also has $\dot{z}>0$ at large times. The
fact that the energy density concerned is QI-compatible (see
Sect.~\ref{sect:del2d}) shows that some of our conditions could be
relaxed.} As shown by Fulling and Davies~\cite{FD}, the expected energy
density in the `in' vacuum $\psi_{\rm in}$
observed along the worldline $(t,0)$ is given by
\begin{equation}
\rho_{\psi_{\rm in}}(t) = 
\frac{1}{12\pi} (p')^{1/2}(t)\left[(p')^{-1/2}\right]''(t),
\end{equation}
where $p(u)=2\tau_u-u$ and $\tau_u$ is defined implicitly by
$\tau_u-z(\tau_u)=u$.\footnote{The condition $\sup |\dot{z}|<1$ 
guarantees the existence of $\tau_u$.} The function $p'$ is a Doppler
shift factor; indeed, writing $\varphi(t)=p'(t)^{-1/2}$, we have
\begin{equation}
\varphi(t) = \sqrt{\frac{1-z'(\tau_t)}{1+z'(\tau_t)}},
\end{equation}
which is smooth, positive and bounded both from above and away from
zero. It follows that $\rho_{\psi_{\rm in}}=(12\pi)^{-1}\varphi''/\varphi$ is
smooth and compactly supported (and thus belongs to $\WW_1$). 

To show that $\rho_{\psi_{\rm in}}$ is QI-compatible, we must show that
the operator $H=-d^2/dt^2+6\pi\rho_{\psi_{\rm in}}$ on $L^2(\RR)$ is
positive. The key observation is that $\rho_{\psi_{\rm in}}$ may be
expressed in terms of the superpotential $U=\varphi'/\varphi$ as
\begin{equation}
\rho_{\psi_{\rm in}} = \frac{1}{12\pi}\left(U'+U^2\right).
\end{equation}
Accordingly $H$
may be written in the manifestly positive form
\begin{equation}
H=-\frac{1}{2}\frac{d^2}{dt^2} + A^*A,
\label{eq:Hident}
\end{equation}
where 
\begin{equation}
A = \frac{1}{\sqrt{2}}\left(\frac{d}{dt} - U\right).
\label{eq:Adef}
\end{equation}
Thus QI-compatibility is (formally) established. 

The above argument is easily made rigorous. Since $\rho_{\psi_{\rm in}}$ 
is smooth, 
the definition of $H$ as a sum of forms is equivalent to the ordinary
operator sum on $D(H)=W^{2,2}(\RR)$. The operator $A$ is defined to be 
the closure of the differential operator~(\ref{eq:Adef}) on
$\CoinX{\RR}$, and it is easy to verify that $A$ and $A^*$ act as
$\pm D-U$ on their common domain $W^{1,2}(\RR)$. Straightforward
calculation shows that $A D(H)\subset D(A^*)$ and that 
the identity~(\ref{eq:Hident}) holds on $D(H)$. The operator $H$ is
then clearly positive and QI-compatibility follows by
Theorem~3.2. Clearly these arguments may also be extended
to trajectories which are only $C^k$ for suitable $k$; we will not
pursue this here. 

We also note that this argument may be run backwards to show that
all energy densities originating from moving mirrors, or from the
class of states considered by Ford~\cite{Ford78}, obey a stronger
quantum inequality than that derived by Flanagan~\cite{Flan}; the
constant $6\pi$ in~(\ref{eq:QI}) 
may be replaced by $12\pi$ for such states.

\section{$\delta$-function pulses in two dimensions} \label{sect:del2d}

We now illustrate how the results of Sect.~\ref{sect:techres} provide
bounds relevant to the quantum interest problem for $\delta$-function
loans in two and four dimensions. By solving the appropriate eigenvalue
problems we will obtain maximal pulse separations for such a loan, and
the minimum quantum interest rates applying to different forms of the
repayment. 

\subsection{Pulse separation}

Consider a loan of the form $\rho(t)=-B\delta(t)$ ($B>0$) in two
spacetime dimensions ($m=1$). Note that $\rho\in\WW_1$. 
This energy density is easily seen not to be QI-compatible on $\RR$ (as
one would expect) by Theorem~3.2 and the fact that the
eigenvalue problem
\begin{equation}
-g'' -6\pi B\delta(t)g(t) = E g(t)
\label{eq:del2d}
\end{equation}
has a solution $g(t) = e^{-3\pi B|t|}$ in $W^{1,2}(\RR)$ with negative
eigenvalue $E=-(3\pi B)^2$. Accordingly, this energy density must be
counterbalanced by later repayments or earlier `deposits' and we may ask
how far these positive energy contributions may be separated from it.
Theorem~3.3 guarantees the existence of at least one 
maximal interval containing the origin on which $\rho$ is QI-compatible 
(provided $\rho$ is QI-compatible on {\em some} such interval) but does
not imply uniqueness. To illustrate this point, we will consider two
maximal separation problems. First, we find the maximum $T$
such that $\rho$ is QI-compatible on $(-T,T)$; second, we repeat the
analysis for intervals of the form $(-\infty,T)$. The second case is
more pertinent to the results of~\cite{FRqi}, where it was shown that
repayments must begin before $T_{\rm FR} = \pi/(24 B)\approx 0.131/B$.
As we will see, the eigenvalue approach will provide sharper bounds. 

By Theorem~3.2, the solution to our first problem is the
maximum $T$ for which the eigenvalue problem~(\ref{eq:del2d}) has no
solutions $g\in W_0^{1,2}(-T,T)$ with $E<0$. Now the discussion
in Sect.~\ref{sect:techres} implies that any solution
to~(\ref{eq:del2d}) must vanish at $t=\pm T$ and be smooth everywhere
except at $t=0$, where it is continuous and satisfies the jump condition
$[g']_{t=0}=-6\pi Bg(0)$. These conditions fix $g$ up to normalization
as $g(t) = \sinh k(T-|t|)$, and impose the condition
\begin{equation}
kT \coth kT= 3\pi BT 
\end{equation}
on $k=\sqrt{-E}>0$. Since the left-hand side is
an increasing function on $\RR^+$ and 
tends to unity as $k\to 0$, we conclude that eigenfunctions with
negative eigenvalues exist if and only if $3\pi BT> 1$. Thus the maximal
separation here is 
\begin{equation}
T_{\rm symm} =\frac{1}{3\pi B}\approx \frac{0.106}{B},
\end{equation}
which represents a tighter bound than $T_{\rm FR}$.
The significance of $T_{\rm symm}$ is that if $T>T_{\rm symm}$ then,
no matter what positive energy `deposits' are made before $t=-T$ or
after $t=+T$, the loan of $-B\delta(t)$ at $t=0$ is forbidden. 

In fact this bound may also be obtained~\cite{Dan} by 
applying the analysis of~\cite{FRqi} to the sampling function 
\begin{equation}
f_{\tau}(t) = \left\{\begin{array}{cl} 
\frac{3}{2\tau}(1-|t|/\tau)^2 & |t|<\tau; \\ 0 & |t|\ge \tau.
\end{array}\right.
\end{equation} 
$T_{\rm symm}$ is then the maximum value
of $\tau$ for which the inequality~(\ref{eq:QI}) holds with
$\rho_\psi(t)=-B\delta(t)$. The function $f_\tau$
is a legitimate sampling function because its square root 
has a weak derivative in $L^2$ and so belongs to
$W^{1,2}(\RR)$, despite not being differentiable at $t=0$
and $t=\pm \tau$. 

The second problem is to seek the maximum $T$ so that $-B\delta(t)$ is
QI-compatible on $(-\infty,T)$. Solving the eigenvalue
problem~(\ref{eq:del2d}) for $g\in W_0^{1,2}(-\infty,T)$ we find
\begin{equation}
g(t) = \left\{ \begin{array}{cl} e^{kt}\sinh kT & t<0; \\ 
\sinh k(T-t) & 0<t<T
\end{array}\right.
\end{equation}
up to normalization, where $k=\sqrt{-E}>0$ as before.
Here we have used the fact that $g$ must be
square integrable and also continuous at $t=0$. Applying the jump
condition gives
\begin{equation}
6\pi BT = kT(1+\coth kT),
\end{equation}
which fails to have solutions with $k>0$ if and only if
\begin{equation}
T\le T_{-\infty}:= \frac{1}{6\pi B} \approx \frac{0.0531}{B}.
\end{equation}
This is the quantity most relevant to the problem studied by Ford and
Roman~\cite{FRqi}: if the energy density vanishes for all $t<0$ and a 
negative $\delta$-loan is made at $t=0$, repayments must begin before
$t=T_{-\infty}$.

In the terms of Theorem~3.3, both 
$(-T_{\rm symm},T_{\rm symm})$ 
and $(-\infty,T_{-\infty})$ are maximal elements of $\OO_\rho$. The fact 
that $T_{-\infty}<T_{\rm symm}$ is of course necessary for consistency. 

\subsection{Quantum interest} \label{sect:d2qi}

Let us now consider what happens if a $\delta$-function compensating
pulse arrives at time $T\le T_{-\infty}$. For what values of the
`interest rate' $\epsilon$ 
is the combined energy density $B[-\delta(t)+(1+\epsilon)\delta(t-T)]$
QI-compatible on $\RR$? 

By Theorem~3.2 we require there to be no solutions in
$g\in W_0^{1,2}(\RR)$ to 
\begin{equation}
-g''(t) +[-\lambda \delta(t) + \mu\delta(t-T)]g(t) = -k^2 g(t)
\end{equation}
where $\lambda = 6\pi B$, $\mu=6\pi(1+\epsilon)B$ and we take $k>0$
without loss of generality. Any solution would be square integrable and 
smooth except at $t=0,T$ where it would be continuous and obey the
jump conditions $[g']_{t=0} =-\lambda g(0)$ and $[g']_{t=T} = \mu g(T)$.
These requirements fix $g$ up to normalization as
\begin{equation}
g(t) = \left\{\begin{array}{cl}
e^{kt} & t\le 0; \\
\cosh kt + \left(1-\lambda/k\right)\sinh kt & 0<t\le T; \\
\left[\cosh kT + \left(1-\lambda/k\right)\sinh kT\right]e^{-k(t-T)}
& T<t,
\end{array}\right.
\end{equation}
and impose the condition $f_{\lambda T,\mu T}(kT) = 0$ on $k$, where
\begin{equation}
f_{\beta,\gamma}(x) =(2x-\beta+\gamma)\cosh x +
\left(2x+\gamma-\beta -\frac{\beta\gamma}{x}\right)\sinh x. 
\end{equation}

Accordingly, $\rho$ is QI-compatible on $\RR$
if and only if $f_{\beta,\gamma}(x)$ is nonzero for all real $x>0$,
which is equivalent to the conditions
\begin{equation}
0<\beta<1 \qquad{\rm and}\qquad \gamma\ge \beta/(1-\beta),
\label{eq:2ddc}
\end{equation}
as we now show. Sufficiency holds because 
each coefficient in the power series
\begin{eqnarray}
f_{\beta,\gamma}(x)&=& \gamma-\beta-\beta\gamma +
\sum_{n=1}^\infty x^{2n}\left(
\frac{\gamma-\beta}{(2n)!} -\frac{\beta\gamma}{(2n+1)!}+
\frac{2}{(2n-1)!}\right) \nonumber\\
&&
+\sum_{n=0}^\infty
x^{2n+1}\left(\frac{2}{(2n)!}+\frac{\gamma-\beta}{(2n+1)!}
\right)
\end{eqnarray}
is positive under these conditions. On the other hand,  if
$f_{\beta,\gamma}$ is nonvanishing for $x>0$, then the intermediate
value theorem implies $f_{\beta,\gamma}(0)\ge 0$ since $f_{\beta,\gamma}$
is everywhere smooth and is positive for large positive $x$. Thus
$\gamma(1-\beta)\ge\beta $, and the required result follows as 
we also have $0<\beta\le 1$ from the condition $0<T\le T_{-\infty}$.

In terms of the original parameters, condition~(\ref{eq:2ddc}) reads
\begin{equation}
0< BT < \frac{1}{6\pi} \qquad{\rm and}\qquad
\epsilon\ge \frac{6\pi BT}{1-6\pi BT}.
\end{equation}
Note that the minimum interest rate $6\pi BT/(1-6\pi BT)$ grows
unboundedly as the duration of the loan approaches its maximum. 
This is a stronger result than that of Ford and
Roman~\cite{FRqi}, whose lower bound on the interest rate remains finite
as the duration of the loan approaches $T_{FR}$. 

Finally, we note that the particular moving mirror trajectory considered
in~\cite{FRqi} has (in our notation) $BT<1/(12\pi)$ and
\begin{equation}
\epsilon = \frac{24\pi BT[72\pi^2 (BT)^2-15\pi BT+1]}{(1-12\pi BT)^3},
\end{equation}
which easily satisfies the conditions above. The
corresponding energy density is therefore QI-compatible on $\RR$, in
accord with the results of Sect.~\ref{sect:mirror} (although the
trajectory itself does not satisfy the conditions imposed there). This
example also exhibits divergent quantum interest rates, but the
divergence here occurs before the duration reaches $T_{-\infty}$.

\section{$\delta$-function pulses in four dimensions} \label{sect:del4d}

\subsection{Pulse separation} \label{sect:d4psep}

We compute maximum pulse separations for the distribution
$\rho=-B\delta(t)$ in four-dimensional massless field theory ($\rho\in\WW_2$). 
First, the maximal symmetric interval $(-T,T)$ on which 
$\rho$ is QI-compatible is found as follows. The appropriate
eigenvalue problem is
\begin{equation}
g''''(t) - 16\pi^2B\delta(t) g(t) = -4k^4 g(t)
\end{equation}
which is to be solved for $k>0$ and $g\in W_0^{2,2}(-T,T)$. Both
$g$ and $g'$ vanish at the endpoints, 
and $g$ is smooth except at $t=0$, where it is $C^2$ with a
discontinuity in $g'''$ determined by the coefficient of the
$\delta$-function. The boundary conditions and continuity of $g$, $g'$ and
$g''$ at $t=0$ fix $g$ up to normalization as
\begin{equation}
g(t) = 2\Psi(kT)\Psi\left(k(T-|t|)\right)
+\Phi^+(kT)\Phi^-\left(k(T-|t|)\right) 
\end{equation}
where
\begin{eqnarray}
\Psi(x) &=& \sinh x\sin x,  \nonumber\\
\Phi^\pm(x) &=& \sinh x\cos x \pm \cosh x\sin x,
\end{eqnarray}
so the discontinuity in $g'''$ at $t=0$ is 
\begin{equation}
[g''']_{t=0} = 4k^3(\sin 2kT +\sinh 2kT).
\end{equation}
Accordingly, the eigenvalue condition $[g''']_{t=0}=16\pi^2 B g(0)$ is
\begin{equation}
f(2kT)=16\pi^2 B T^3
\end{equation}
where
\begin{equation}
f(x)= x^3\frac{\sinh x+\sin x}{\cosh x+\cos x -2}.
\end{equation}
Now $f$ is increasing on $(0,\infty)$ and tends to $24$ as $x\to 0^+$,
so the eigenvalue equation fails to have solutions if and only if 
$16\pi^2 BT^3\le 24$. Thus $T_{\rm symm}=(3/(2\pi^2 B))^{1/3}$. 

Repeating the analysis for QI-compatibility on $(-\infty, T)$, we 
find eigenfunctions as follows:
\begin{equation}
g(t) = \left\{\begin{array}{cl} 
2\sin kT \Psi(k(t-T)) 
-(\sin kT+\cos kT)\Phi^-(k(t-T)) & t>0; \\
e^{k(t-T)} (C\sin kt-D\cos kt) & t<0,
\end{array}\right.
\end{equation}
where $C = \frac{1}{2}(\sin 2kT+\cos 2kT-e^{2kT})$, $D=
\frac{1}{2}(\sin 2kT-\cos 2kT-e^{2kT}+2)$. The jump condition becomes
$f(2kT) = 16\pi^2 BT^3$ where
\begin{equation}
f(x) = \frac{x^3}{1+e^{-x}(\cos x-\sin x -2)}.
\end{equation}
This function obeys $f(x)\ge 3$ for $x\ge 0$ with equality at $x=0$. 
Thus the maximal pulse separation here is
\begin{equation}
T_\infty = \left(\frac{3}{16\pi^2 B}\right)^{1/3}\approx 0.267B^{-1/3}
\end{equation}
which improves on Ford and Roman's upper bound $0.338
B^{-1/3}$~\cite{FRqi}. We emphasize that this bound strengthens 
(and does not contradict) their
result.\footnote{Incidentally, it turns out that the square root of the 
sampling function employed in~\cite{FRqi} is not an element of
$W^{2,2}(\RR)$ and would not be a legitimate sampling function in our
framework. Nonetheless, it is certainly true that their bound is
a valid upper bound for the pulse separation. This issue does not arise
in two dimensions because the square root of the sampling function
of~\cite{FRqi} is an element of $W^{1,2}(\RR)$.}
Our assertions concerning $f$ are proved by the following means: let
$q(x)$ be the denominator, then $q(0)=q'(0)=q''(0)=0$ and $q'''(x)=
(2-4\sin x)e^{-x}\le 2$ for all $x\in\RR^+$, so
Taylor's theorem with remainder shows
\begin{equation}
q(x) \le \frac{x^3}{3}
\end{equation}
on $\RR^+$. Since $q\ge 0$ on $\RR^+$, we have $3\le f(x)$ 
with equality in the limit $x\to 0$.

\subsection{Quantum interest}

Repeating the analysis of Sect.~\ref{sect:d2qi}, we consider the
QI-compatibility of $\rho(t) = B[-\delta(t)+(1+\epsilon)\delta(t-T)]$ by
seeking solutions $g\in W^{2,2}_0(\RR)$ to the equation 
\begin{equation}
g''''(t)+[-\lambda\delta(t)+\mu\delta(t-T)]g(t)=-4k^4g(t)
\end{equation}
where $\lambda=16\pi^2B$, $\mu=16\pi^2B(1+\epsilon)$, and $k>0$. 
On $(-\infty,T)$, any solution must take the form
\begin{equation}
g(t) = e^{kt}(C\cos kt+D\sin kt)-C \theta(t)\frac{\lambda}{4k^3}\Phi^-(kt)
\label{eq:gform}
\end{equation}
while on $(T,\infty)$ we have
\begin{equation}
g(t) = e^{-kt}(E\cos kt+F\sin kt).
\end{equation}
The matching conditions at $t=T$---continuity
of $g$, $g'$, and $g''$, and a discontinuity $[g''']_{t=T}=-\mu g(T)$ in
$g'''$---fix $C,D,E,F$ up to overall scale as
\begin{eqnarray}
C &=&  8k^3(\cos kT+\sin kT), \nonumber\\
D &=& (8k^3-2\lambda)\sin kT-8k^3\cos kT, \nonumber\\
E &=& [(8k^3-\lambda)e^{2kT}+\lambda]\cos kT+
[(\lambda-8k^3)e^{2kT}+\lambda]\sin kT, \nonumber \\
F &=& [(8k^3 -\lambda)e^{2kT}+\lambda](\sin kT+\cos kT),                    
\end{eqnarray}
and impose the condition $f_{\lambda T^3,\mu T^3}(2kT)=0$ on $k$, where
\begin{equation}
f_{\beta,\gamma}(x)=x^6+x^3(\gamma-\beta)+\beta\gamma[e^{-x}(1+\sin
x)-1]=0.
\end{equation}
Since
$f_{\beta,\gamma}(x)\to +\infty$ as $x\to +\infty$ and
$f_{\beta,\gamma}(x) = -\frac{1}{2}\beta\gamma x^2+O(x^3)$ for small
$x$, we see that this condition has a solution for any positive values
of $\lambda$ and $\mu$. 

We conclude that for any $B>0$ there is no `interest rate' $\epsilon$ 
for which the energy density considered here 
can be QI-compatible in four dimensions, and that there is no physical
state with an expected energy density of this form. (Note that one
would not expect to produce such a simple energy density
by moving mirrors in four dimensions; other contributions also appear, at
least in the nonrelativistic approximation~\cite{FV}.)

A legitimate question at this stage is: what does this result tell us
about more physical smooth candidate energy densities?\footnote{We are
grateful to the referee for raising this issue.} In other words,
can one also rule out energy densities consisting of two strongly
peaked, but nonetheless smooth, pulses? We hope to address this in
detail elsewhere and give only a qualitative answer here. Suppose one is
given a sequence $\rho_n$ of smooth, compactly supported functions
which converge to $\rho(t)=B[-\delta(t)+(1+\epsilon)\delta(t-T)]$ for some
values of $B,T,\epsilon$ in the $\|\cdot\|_{-2,\infty}$-norm.\footnote{
Such sequences are easily obtained by mollifying $\rho$, 
cf.\ Theorem~4.1.4 in~\cite{Hor1} and Lemma~3.15 in~\cite{Adams}.}
Then Theorem~B.1 of Appendix~\ref{appx:B} 
shows that the corresponding operators 
$H^{(2)}_n=D^4+c_2\rho_n$ on $L^2(\RR)$ converge to the
limiting operator $H^{(2)}=D^4+c_2\rho$ in the {\em norm resolvent
sense} (see \S VIII.7 in~\cite{RSi}). Now the spectrum of self-adjoint
operators cannot expand in a norm resolvent limit (indeed, the same is
true for the weaker strong resolvent convergence---see Theorem~VIII.24
in~\cite{RSi}) so, since $H^{(2)}$ has at least one negative eigenvalue,
so must $H^{(2)}_n$ for all sufficiently large $n$. Accordingly,
the $\rho_n$ fail to be QI-compatible for all sufficiently large $n$. 

Thus we see that our result does have implications for smooth candidate
energy densities, although the precise quantitative details require
further work. 

\section{$\delta$-function pulse with a positive step in four dimensions}
\label{sect:delstep}

Our last example concerns a negative-energy $\delta$-function pulse at 
$t=0$ followed 
at time $t=T$ by a positive energy density of finite magnitude but
infinite duration
\begin{equation}
\rho(t) = -B\left[\delta(t) +\frac{1+\epsilon}{\tau}\theta(t-T)\right]
\end{equation}
for $B,\epsilon>0$, where $\tau$ is a timescale inserted for dimensional
reasons.  This candidate energy density belongs to
$\WW_2$ because the second term is a positive element of
$L^{\infty}(\RR)$ and hence of $W^{-2,\infty}(\RR)$. 
Note that the total `repayment' of positive energy
is infinite for all values of these parameters; nonetheless we will see
that QI-compatibility forces $\epsilon$ to increase without bound as $T$
approaches the maximal value $T_{-\infty}$ obtained above.  

It will be convenient to introduce dimensionless quantities 
$\beta=16\pi^2BT^3$ and
$\gamma=[64\pi^2 B(1+\epsilon)\tau^{-1}T^4]^{1/4}$. For $t<T$ 
the analysis is identical to that in the preceding section
and the eigenfunction $g$ takes the form~(\ref{eq:gform}) in this range. 
When $T<t$, the eigenvalue problem is 
\begin{equation}
g'''' = -4 k'{}^4g,
\end{equation}
where $k'{}^4=k^4+(\gamma/2T)^4$ 
so the solution takes the general form
\begin{equation}
g(t) =e^{-k't}[E\cos k'(t-T)+F\sin k'(t-T)],
\end{equation}
since $g$ must be square integrable. 
Continuity of $g,g',g''$ at $t=T$ fixes the coefficients
up to scale as
\begin{eqnarray}
C &=& x^3(x+y)\left(x\cos\frac{x}{2}+y\sin\frac{x}{2}\right)e^{x/2},
\nonumber\\ 
D &=& 2\beta\left[y^2\Phi^-(x/2) - x^2\Phi^+(x/2)\right] 
 -x^3(x+y)\left(y\cos\frac{x}{2}-x\sin\frac{x}{2}\right)e^{x/2},
\nonumber\\
E &=& \frac{Ce^{y/2}}{x^3}\left[x^3 e^{x/2}\cos\frac{x}{2}
-2\beta\Phi^-(x/2)
\right] +D e^{(x+y)/2}\sin\frac{x}{2}, \nonumber \\
F &=& \frac{Ce^{y/2}}{xy^2}\left[x^3 e^{x/2}\sin\frac{x}{2}-
2\beta\Phi^+(x/2)
\right] -D\frac{x^2}{y^2} e^{(x+y)/2}\sin\frac{x}{2},
\end{eqnarray}
where $x=2kT$ and $y=2k'T=(x^4+\gamma^4)^{1/4}$.
The final matching requirement, continuity of
$g'''$ at $t=T$, imposes the eigenvalue condition 
$f_{\beta,\gamma}(x)=0$ on $x$ where $f_{\beta,\gamma}$,
defined by
\begin{equation}
f_{\beta,\gamma}(2kT) = \frac{2T^2(k'-k)}{k(k+k')}\left[g'''\right]_{t=T}
\end{equation}
may be shown, after a tedious calculation, to take the form
\begin{eqnarray}
f_{\beta,\gamma}(x)&=&
\gamma^4\left[x^3e^x+
\beta\left(2-\cos x+\sin x-e^x\right)\right]\nonumber \\
&&\qquad+4\beta x(y-x)\left[y{}^2(\cos x-1)-
x y\sin x-x^2\right].
\end{eqnarray}

The QI-compatibility of $\rho$ is equivalent to the absence of zeros of
$f_{\beta,\gamma}(x)$ for $x>0$. Since $f_{\beta,\gamma}(x)\rightarrow 
+\infty$ as $x\rightarrow +\infty$, a necessary condition for
QI-compatibility is
that $f_{\beta,\gamma}(x)$ should be non-negative for
small $x>0$. Now
\begin{equation}
f_{\beta,\gamma}(x)=\gamma\left[\gamma^3-
\beta\left(4+4\gamma+2\gamma^2
+\frac{1}{3}\gamma^3\right)\right]x^3+O(x^4),
\end{equation}
so our necessary condition becomes
\begin{equation}
\label{eq:cubic}
P(\gamma):=\gamma^3-\beta\left(4+4\gamma+2\gamma^2
+\frac{1}{3}\gamma^3\right)\ge 0.
\end{equation}
In addition, 
the pulse separation bound of Sect.~\ref{sect:d4psep} entails that
$0<\beta\le 3$. It is now easy to see that
condition~(\ref{eq:cubic}) implies
\begin{equation}
0<\beta<3 \qquad{\rm and}\qquad \gamma\ge \gamma_0 \label{eq:QIcond}
\end{equation}
as necessary conditions for QI-compatibility. Here,
$\gamma_0$ is the unique real root of $P(\gamma)$, given by
\begin{equation}
\gamma_0 = \frac{2\beta}{3-\beta} + \frac{4\sqrt{3\beta}}{3-\beta} 
\cosh\left[\frac{1}{3}
\cosh^{-1}\left(\frac{\beta^2+27}{12\sqrt{3\beta}}\right)\right].
\end{equation}
Note that $\gamma_0$ increases monotonically and without bound 
as the maximal loan term $\beta=3$ is approached. 

In fact, the conditions~(\ref{eq:QIcond}) are not only necessary, but
also sufficient for QI-compatibility as we now show. To do this, we must
prove that $f_{\beta,\gamma}(x)$ does not vanish for $x>0$
if~(\ref{eq:QIcond}) holds. 
Now, the analysis at the end of Sect.~\ref{sect:d4psep} concerning the
function $f$ implies that 
\begin{equation}
2-\cos x+\sin x-e^x \ge -\frac{x^3}{3}e^x.
\end{equation}
Using this inequality, along with condition~(\ref{eq:cubic}) in the form
\begin{equation}
\gamma^4\ge \beta\gamma\left(4+4\gamma+2\gamma^2
+\frac{1}{3}\gamma^3\right),
\end{equation}
we have 
\begin{eqnarray}
f_{\beta,\gamma}(x)
&\ge & 2\beta\gamma(\gamma^2+2\gamma+2)x^3e^x 
+4\beta x(y-x)\left[y{}^2(\cos x-1)-
x y\sin x-x^2\right] \nonumber\\
&=&\kappa 
\left[\left(\frac{1}{2}\gamma^2+\gamma+1\right)(y+x)
(y{}^2+x^2)\,e^x
+\gamma^3\left(y{}^2\,\frac{\cos x-1}{x^2}
-y\,\frac{\sin x}{x}-1\right)\right] \nonumber \\
&&
\end{eqnarray}
where 
\begin{equation}
\kappa = \frac{4\beta\gamma x^3}{(y+x)(y{}^2+x^2)}
\end{equation}
is positive. 

Next, applying the elementary bounds $(\sin x)/x\leq1$ 
and $(\cos x-1)/x^2\ge -1/2$ for $x>0$, we obtain 
\begin{equation}
f_{\beta,\gamma}(x)
\ge \kappa'\left[(y+x)
\left(1+\frac{x^2}{y{}^2}\right)\,e^x 
-\frac{\gamma}{\frac{1}{2}\gamma^2+\gamma+1}
\left(\frac{\gamma^2}{2}+\frac{\gamma^2}{y}+\frac{\gamma^2}{y{}^2}
\right)\right] 
\end{equation}
where
\begin{equation}
\kappa' = \kappa y{}^2
\left(\frac{1}{2}\gamma^2+\gamma+1\right)>0.
\end{equation}
Finally, we have $y>\gamma$, so 
the first term in the braces is greater than $\gamma$ [because,
additionally, $x>0$ and $e^x>1$] while the second term is less than
$\gamma$. Hence $f_{\beta,\gamma}(x)\geq0$ for all
$x>0$ as required. 

To summarize, we restate the QI-compatibility conditions~(\ref{eq:QIcond})
in terms of
the original parameters $B$, $T$, and $\epsilon$, in which form they read
\begin{equation}
0<BT^3<\frac{3}{16\pi^2} \qquad {\rm and} \qquad B(1+\epsilon)\ge
\frac{\gamma_0^4\tau}{64\pi^2 T^4}.
\end{equation}
With $B$ fixed, it is clear that $\epsilon\to\infty$ as $T$ approaches
the maximal value $T_{-\infty}$, because $\gamma_0\to\infty$ as
$\beta\to 3$.

\section{Conclusion}

We have presented a new viewpoint on quantum inequalities,
in terms of eigenvalue problems, and have demonstrated its utility with
reference to the circle of ideas surrounding the quantum interest
conjecture. In particular, we have obtained optimal bounds on the pulse
separations and quantum interest rates which improve on the values found
in~\cite{FRqi} and indeed rule out the double $\delta$-function pulse model
in four dimensions. Our general approach has also given a simple 
confirmation that moving mirrors in two dimensions yield QI-compatible
energy densities for a large class of trajectories.
We conclude with a few remarks. 

First, we have seen that quite singular candidate energy densities, in
particular (derivatives of) $\delta$-functions, may
be accommodated within our framework. Although the classes $\WW_m$
studied in Sect.~\ref{sect:techres} are more general than strictly
required to treat the examples studied here,
they strongly suggest the possibility of extending the results
of~\cite{Fqi} to (certain classes of) non-Hadamard states. 

Second, it is noteworthy that sampling functions, which were the
key element allowing Ford and coworkers to progress beyond the 
pointwise unboundedness of the energy density, are almost eliminated
from the present treatment. The eigenvalue viewpoint may thus be
considered as a `coordinate-free' version of the quantum inequalities.
Of course, there will be many circumstances in which the 
eigenvalue problem cannot be solved analytically. One would then need to
fall back on the use of sampling functions much as in~\cite{FRqi,Pret}.
However, the eigenvalue viewpoint can still be of use, as the strongest
information regarding QI-compatibility will be obtained from
sampling functions of the form $|g(t)|^2$, where $g$ is chosen to be a
good approximation to an eigenfunction e.g., by WKB methods. 

Third, our viewpoint provides a more intuitive understanding 
of the quantum interest conjecture using the analogy with
quantum mechanics for a particle moving on a line. This is a precise
correspondence for two-dimensional quantum fields, but much of the
intuition is also valid in the four-dimensional case. A negative energy loan 
may be considered as a potential well in which bound states would 
form in the absence of a suitable positive energy repayment, which
acts as a barrier. To repay the loan, the barrier must disrupt the
tail of the bound state wavefunction, essentially bouncing the particle
out of the potential well. Since the tail decays
rapidly away from the well, it is not surprising that the repayment must
be greater than the original loan in some sense, and that it must become
larger as the term of the loan is increased. In a precise sense {\em no\/}
positive barrier
can have a greater effect than a perfectly reflecting wall. The maximum
pulse separation is therefore the minimum delay between the well and a 
wall such that bound states may still form. In the examples studied
here, the interest rate diverges as the maximum loan term is approached.
We believe this to be very natural: given the existence of maximal pulse
separations there must be a physical mechanism which prevents one from exceeding
this bound, and it seems more natural that the interest rate should
diverge as one approaches the limit rather than for such a mechanism to
switch on suddenly at the maximal separation. We intend to return to this 
issue elsewhere.  

To give a direct physical interpretation of our results in
terms of quantum field theory, consider the problem of verifying
that a given candidate energy density is the energy density of a quantum
state. It would be necessary to construct a two-point function with normal
ordered energy density agreeing with the candidate on a specified worldline. In
effect, one would be attempting to solve the Klein-Gordon equation subject to
data specified on this worldline, for a bisolution obeying the 
positivity conditions necessary for it to arise from a state (and which,
in turn, imply the quantum inequalities~\cite{Fqi}). This is a somewhat
involved process and it is not easy to gain direct insight into the
class of functions for which it is possible. The results obtained here and
in~\cite{FRqi,Pret} suggest very strongly that these conditions are
highly restrictive and also function in a semi-local fashion: if the
bisolution develops a region in which it fails to obey positivity, this
cannot be repaired by modifying the candidate energy density at distant
times.  

Finally, in terms of practical applications our approach is currently  
limited to massless fields in even dimensions (see, however, the
comments at the end of the introduction). Nonetheless the general
viewpoint is very natural and could well form part of a deeper
understanding of quantum inequalities and their ramifications.

\appendix
\section{Proof of Lemma~3.1}
\label{appx:Pf}

We begin by considering the case $I=\RR$. To simplify
the notation, we write $\Am_\RR$ as $\Am$ etc. 

\medskip
{\em Lemma A.1\/}.
Suppose $\rho\in W^{-m,\infty}(\RR)$ (respectively,
$\rho\in W^{-m,2}(\RR)\cap W^{-m,\infty}(\RR)+
\left(W^{-m,\infty}(\RR)\right)_\epsilon$) is real. 
Then its associated
quadratic form is $\Am$-form bounded (respectively, $\Am$-form compact). 
\medskip

{\em Proof.\/} Let $\rho$ be an arbitrary element of
$W^{-m,\infty}(\RR)$. We have
\begin{equation}
|\rho(f,f)|=|\act{|f|^2}{\rho}| \le \|\rho\|_{-m,\infty}
\|\,|f|^2\,\|_{m,1}
\end{equation}
by definition of $\|\cdot\|_{-m,\infty}$. Now (by Theorem~4.13 
in~\cite{Adams}, Leibniz' rule, and the Cauchy-Schwarz
inequality) there exists a $C>0$ such that
\begin{equation}
\|\,|f|^2\,\|_{m,1}\le C\left(\|D^mf\|_2^2+\|f\|_2^2\right) =
C\|f\|_{+1}^2, \qquad \forall f\in W^{m,2}(\RR),
\end{equation}
so we have
\begin{equation}
|\rho(f,f)|\le C \|\rho\|_{-m,\infty}\|f\|_{+1}^2.
\end{equation}
This has two consequences. First, 
$\rho$ is $\Am$-form bounded, which proves the first part of the
statement. Second, 
the self-adjoint operator $R$ on $\HH_{+1}$ associated with $\rho$ by
\begin{equation}
\ip{f}{Rg}_{+1} = \rho(f,g), \qquad\forall f,g\in\HH_{+1} \label{eq:Rdef}
\end{equation}
has operator norm $\|R\|\le C\|\rho\|_{-m,\infty}$. Accordingly, if
$\rho_n\to\rho$ in $W^{-m,\infty}(\RR)$, their corresponding operators
obey $R_n\to R$ in norm. 

In particular, if $\rho\in W^{-m,2}(\RR)\cap W^{-m,\infty}(\RR)+
\left(W^{-m,\infty}(\RR)\right)_\epsilon$, its corresponding operator
may be regarded as a norm limit of a sequence of operators derived from
distributions in $W^{-m,2}(\RR)\cap W^{-m,\infty}(\RR)$. We will see
presently that such operators are Hilbert-Schmidt; accordingly their
norm limit is compact, thus proving that $\rho$ is $\Am$-form compact. 

To complete the proof, then, suppose 
$\rho\in W^{-m,2}(\RR)\cap W^{-m,\infty}(\RR)$. In
particular (cf.~Theorem~7.63 in~\cite{Adams}) 
$\rho$ is a tempered distribution whose Fourier transform
$\widehat{\rho}$ is a measurable function obeying
\begin{equation}
\int \frac{|\widehat{\rho}(k)|^2}{1+k^{2m}}\,dk <\infty.
\label{eq:rho}
\end{equation}
Let $f,g$ be Schwartz test functions. We have 
\begin{equation}
\ip{g}{Rf}_{+1} = \act{g\overline{f}}{\rho} = 
\int \frac{dk}{2\pi}\, \overline{\widehat{g\overline{f}}(k)}\widehat{\rho}(k)
\end{equation}
by Parseval's theorem, so
\begin{equation}
\ip{g}{Rf}_{+1} =\int \frac{dk}{2\pi}\,\int \frac{dk'}{2\pi}\, 
\overline{\widehat{g}(k')}
\widehat{\rho}(k)\widehat{f}(k'-k).
\end{equation}
We may interchange the order of integration by Fubini's theorem (which is
justified owing to Eq.~(\ref{eq:rho}) and 
the rapid decay of $\widehat{f},\widehat{g}$) and, since the resulting
expression holds for arbitrary $g$, identify
\begin{equation}
(1+{k'}^{2m})\widehat{Rf}(k') = \int \frac{dk}{2\pi}\,
\widehat{\rho}(k)\widehat{f}(k'-k) =
\int \frac{dk}{2\pi}\,
\widehat{\rho}(k'-k)\widehat{f}(k)
\label{eq:Rint}
\end{equation}
for almost all $k'$. 

It remains to observe that, by Fubini's theorem,
\begin{equation}
\int dk\,dk'\, \frac{|\widehat{\rho}(k-k')|^2}{(
1+k^{2m})(1+{k'}^{2m})} =
\int du\, \frac{|\widehat{\rho}(u)|^2}{1+u^{2m}}
\int dk\,\frac{1+u^{2m}}{(1+k^{2m})[1+(u-k)^{2m}]}
\end{equation}
and that the $k$-integral defines a bounded function of $u$, to
establish that $R$ agrees with a
Hilbert-Schmidt operator on Schwartz test functions. Since these
are dense in $\HH_{+1}$ we conclude that $R$ is Hilbert-Schmidt, as
required. $\square$
\medskip

{\em Proof of Lemma~3.1\/}. Let $\HH_{+1}$ and
$\HH_{+1,I}$ be the form domains $Q(\Am)=W^{m,2}(\RR)$ and
$Q(\Am_I)=W_0^{m,2}(I)$ equipped with the
appropriate form inner products. It is easy to show that
\begin{equation}
R_I = \jmath^* R\jmath
\end{equation}
where $R$ and $R_I$ are the self-adjoint
operators on $\HH_{+1}$ and $\HH_{+1,I}$
associated with $\rho_\RR$ and $\rho_I$ respectively
and 
$\jmath:\HH_{+1,I}\to \HH_{+1}$ is the isometry induced by the embedding of 
$W_0^{m,2}(I)$ in $W^{m,2}(\RR)$ described below Eq.~(\ref{eq:rhoIdef}).

It follows that $R_I$ inherits any of the properties of boundedness,
compactness, or positivity possessed by $R$.
Thus Lemma~3.1 follows from Lemma~A.1. $\square$
\medskip

\section{A convergence result} \label{appx:B}

{\em Theorem B.1\/}.
Suppose $\rho\in\WW_m$ is the limit in the
$\|\cdot\|_{-m,\infty}$-norm of a sequence $\rho_n\in\WW_m$. Let 
$\Hm$ and
$\Hm_{n}$ be the self-adjoint operators on $L^2(\RR)$ 
constructed by Lemma~3.1
and Theorem~2.4 applied to $\rho$ and $\rho_n$ respectively. 
Then $\Hm_{n}\to \Hm$ in the norm resolvent sense. 
\medskip

{\em Proof.} In the notation of Lemma~A.1 and
by the comments following Eq.~(\ref{eq:Rdef}), the operators $R_n\to R$
in norm on $\HH_{+1}$. Setting $F=(\Am+\openone)^{-1/2}$ this entails 
$F\Hm_{n}F\to F\Hm F$ 
in the operator norm on $L^2(\RR)$ and also implies that the
$\Hm_n$ and $\Hm$ have a common lower bound, say $-M$. The result now
follows (cf. Theorem~VIII.25 in~\cite{RSi}) by the calculation
\begin{eqnarray}
(\Hm_n+t\openone)^{-1}
& =& F\left[t\openone + F
\left(\Hm_n-t\Am\right)F\right]^{-1}F
\nonumber\\
&\longrightarrow & 
F\left[t\openone + F
\left(\Hm-t\Am\right)F\right]^{-1}F
\nonumber\\
&=& (\Hm+t\openone)^{-1}
\end{eqnarray}
for any $t>|M|$, with convergence in operator norm on
$L^2(\RR)$. $\square$

\end{document}